\documentclass{Interspeech}

\interspeechcameraready

\title{
PersonaTAB: Predicting Personality Traits using Textual, Acoustic, and Behavioral Cues in Fully-Duplex Speech Dialogs
}

\author[affiliation={1}]{Sho}{Inoue}
\author[affiliation={2,1}]{Shuai}{Wang}
\author[affiliation={1,3}]{Haizhou}{Li}

\affiliation{SRIBD, School of Data Science}{The Chinese University of Hong Kong, Shenzhen}{China}
\affiliation{School of Intelligence Science and Technology}{Nanjing University, Suzhou}{China}
\affiliation{Department of ECE}{National University of Singapore}{Singapore}
\email{shoinoue@link.cuhk.edu.cn}

\usepackage{comment}
\usepackage{footmisc}
\usepackage{stfloats}  

\usepackage{arydshln}
\usepackage[table]{xcolor}

\usepackage{lipsum}

\usepackage[most]{tcolorbox}
\usepackage{float}
\usepackage{xspace}
\tcbset{
  aibox/.style={
    width=474.18663pt,
    top=10pt,
    colback=gray!15,
    colframe=black,
    colbacktitle=black,
    enhanced,
    center,
    attach boxed title to top left={yshift=-0.1in,xshift=0.15in},
    boxed title style={boxrule=0pt,colframe=white,},
  }
}
\newtcolorbox{AIbox}[2][]{aibox,title=#2,#1}

\tcbset{
  aiboxsmall/.style={
    width=210pt,
    top=8pt,
    bottom=-5pt,
    colback=gray!15,
    colframe=black,
    colbacktitle=black,
    enhanced,
    center,
    attach boxed title to top left={yshift=-0.1in,xshift=0.15in},
    boxed title style={boxrule=0pt,colframe=white,},
  }
}
\newtcolorbox{AIboxSmall}[2][]{aiboxsmall,title=#2,#1}

\begin{document}

\maketitle

\begingroup
  \renewcommand\thefootnote{}
  \footnotetext{%
    Correspondence to Shuai Wang: \texttt{shuaiwang@nju.edu.cn}%
  }%
\endgroup

\begin{figure*}[!t]
\centering
\includegraphics[width=0.95\textwidth]{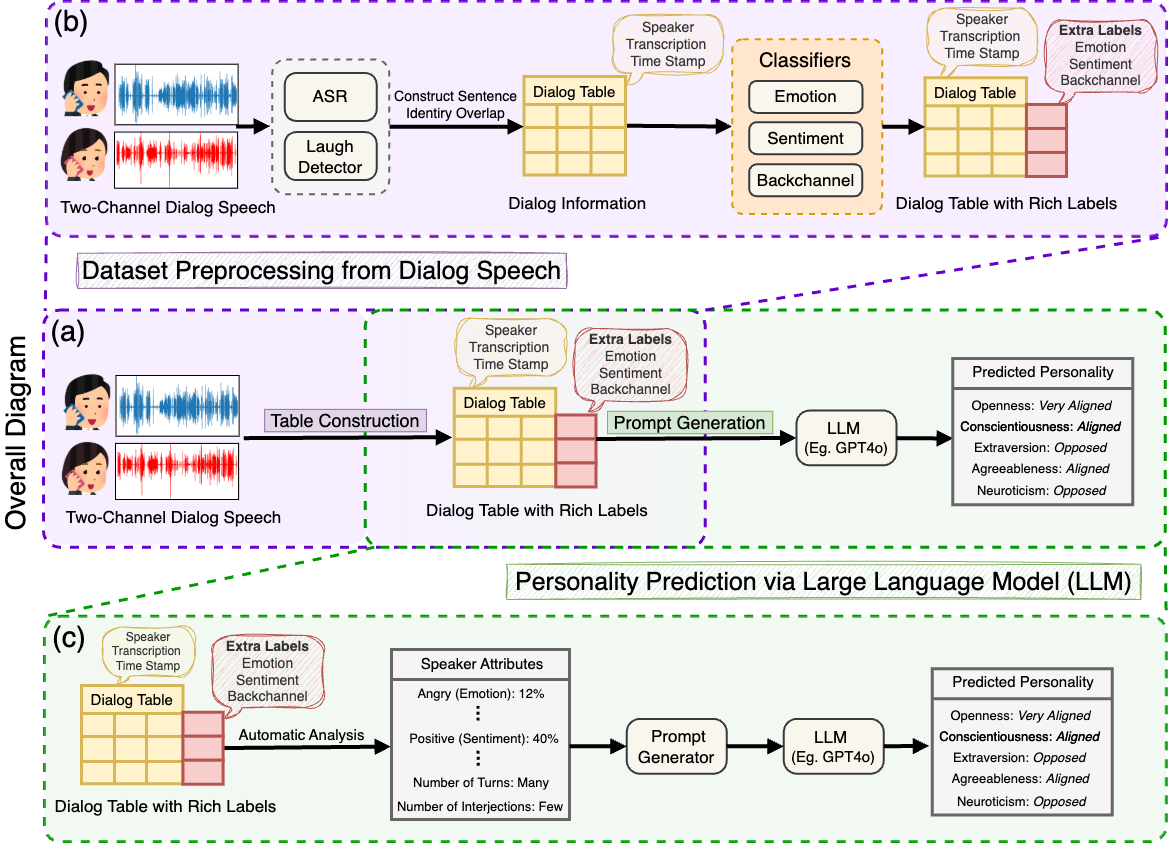}
\caption{
Diagrams of: (a) Overall PersonaTAB Pipeline; (b) Dataset Preprocessing from Two-Channel Speech Dialog Data; (c) Personality Prediction from Speaker Attributes using Large Language Models (LLMs).
}
\label{fig:wide1}
\end{figure*}

\begin{abstract}

Despite significant progress in neural spoken dialog systems, personality-aware conversation agents---capable of adapting behavior based on personalities---remain underexplored due to the absence of personality annotations in speech datasets. We propose a pipeline that preprocesses raw audio recordings to create a dialogue dataset annotated with timestamps, response types, and emotion/sentiment labels.  We employ an automatic speech recognition (ASR) system to extract transcripts and timestamps, then generate conversation-level annotations. Leveraging these annotations, we design a system that employs large language models to predict conversational personality. Human evaluators were engaged to identify conversational characteristics and assign personality labels. Our analysis demonstrates that the proposed system achieves stronger alignment with human judgments compared to existing approaches. 

\end{abstract}

\section{Introduction}


Conversational agents have been extensively studied and deployed in applications such as voice assistants and chatbots. These agents are generally into two models: the turn-taking (streaming) model and the fully-duplex (interaction) model~\cite{SurveySDM}.  In the turn-taking model, the agent responds only after the user completes speaking~\cite{SpeechGPT,QAandSC,Llamaomni}, whereas the fully-duplex model permits simultaneous speech~\cite{dGSLM,Moshi,DialogueScheme}, enabling richer interactions that include interjections and backchanneling.  Notably, backchanneling can stimulate speaker engagement~\cite{Pipek2007OnBI} and influence speech flow~\cite{Bavelas2000,Tolins2015}. We focus on the \emph{fully-duplex setting} due to its closer resemblance to natural conversation.

Despite recent progress, personality-aware conversational agents remain underexplored. 
These agents may exhibit diverse behaviors and attitudes in identical scenarios due to personality traits, influencing both their responses~\cite{Sanfiel2014,Mairesse2007} and their use of interjections~\cite{Park2024,Han2010,1987129}. Current systems often fail to adapt behavior based on personality, instead relying on static or random personas. Conversation datasets with explicit personality labels are limited. To bridge this gap, we propose a method for predicting conversational personality using a fully-duplex speech dialog dataset.


Prior research has primarily employed text-based sources such as social media posts~\cite{PDContinuous,PDmultiplesns,PDChatGPT}, essays~\cite{PDChatGPT,PDAPP,PDbottomtop,PDgraywolf}, and movie scripts~\cite{PDDyadic} for personality prediction. Although~\cite{PDChatGPT} demonstrated that large language models (LLMs) can infer personality traits from text, they have not fully considered both chat history and conversational behaviors.

In this paper, we construct a dialog dataset with personality labels derived exclusively from speech data. We introduce a pipeline that preprocesses raw speech into a structured richly-labeled dialog format. Inspired by~\cite{PDChatGPT}, we develop a prediction framework employing Large Language Models (LLMs) to assign personality labels to speakers. Our approach integrates various aspects of conversational behavior, including textual content (emotion and sentiment), acoustics (laughter), and additional characteristics (e.g. interjections and speaking dominance). This framework enables accurate, context-sensitive personality inference from speech-based dialogues. The contributions of this work are summarized as follows:

\begin{itemize} 
\item We introduce a pipeline that preprocesses raw audio recordings to create a dialogue dataset annotated with timestamps, response types, and emotion/sentiment.
\item We design a system to predict conversational personality from speech dialogues. This study is the first to integrate interjection behaviors, acoustics, and other conversational traits with textual data for personality prediction.
\item We validate both the generated dataset and prediction model through human evaluations, offering valuable resources for developing conversational systems.
\end{itemize}

Supplementary materials---including the implementation, the annotated dataset, sample prompts for LLMs, and sample pages for human survey---are available on our project page\footnote{\textbf{Project Page: }\url{https://github.com/shinshoji01/Personality-Prediction-for-Conversation-Agents}\label{ft:project_page}}.



\begin{figure*}[!ht]
  \includegraphics[width=2.0\columnwidth]{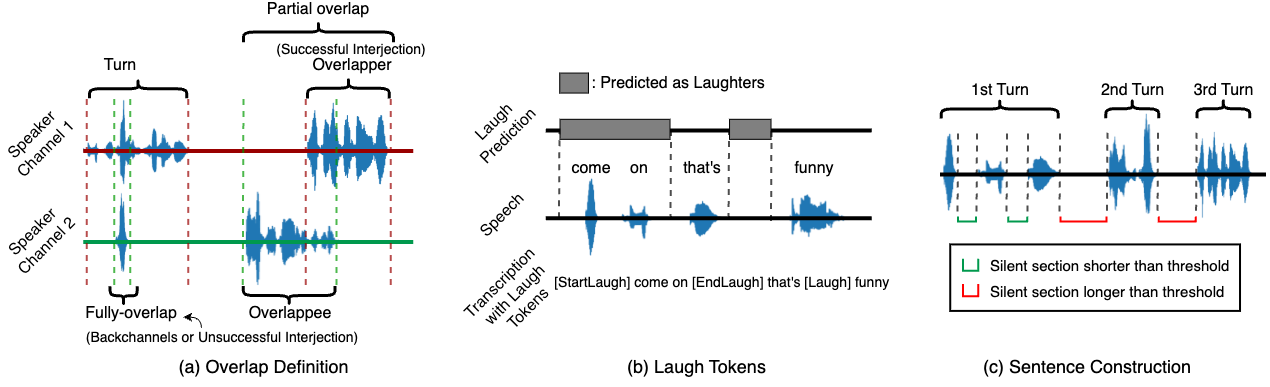}
  \centering
  \caption{
  Visualization of examples of (a) Turns and Overlaps; (b) Laugh Token Integration; (c) Sentence Concatenation from Word-Level Time Stamps.
  }
  \label{fig:overall}
\end{figure*}

\section{Preliminaries}

\subsection{Personality Traits in Conversation}\label{sec:bfi}
We adopt the Big Five Personality Model~\cite{BFI}, a widely recognized psychological framework that identifies patterns in individuals' thoughts, emotions, and behaviors along five dimensions. Each dimension is detailed as follows, with opposite terms in the blankets:

\begin{itemize}
\item \textit{\textbf{openness}} (\textit{closeness}): intellectual, imaginative, independent-minded
\item \textit{\textbf{conscientiousness}} (\textit{lack of direction}): orderly, responsible, dependable
\item \textit{\textbf{extraversion}} (\textit{introversion}): talkative, assertive, energetic
\item \textit{\textbf{agreeableness}} (\textit{antagonism}): good-natured, cooperative, trustful
\item \textit{\textbf{neuroticism}} (\textit{emotional stability}): emotional instability, irritability, anxiety, self-doubt, depression
\end{itemize}

\subsection{Terms in Fully-Duplex Setting}\label{sec:definition}
We define turn-taking events based on prior studies~\cite{TurntakingICLR,Duncan1972,Ameka2006,SKANTZE2021101178}. These events encompass turn-taking behaviors, backchannels, and interjections. Fig.~\ref{fig:overall}(1) illustrates the definitions:

\begin{itemize}
\item \textbf{Turn}: A speaker's contribution from the start of their speech until they stop or another speaker takes over.
\item \textbf{Partial overlap}: An instance where the \textit{overlappee}\footnote{Overlappee: the speaker/response that gets overlapped.\label{ft:overlappee}} stops speaking before the \textit{overlapper}\footnote{Overlapper: the speaker/response that overlaps the speech.\label{ft:overlapper}} finishes speaking.
\item \textbf{Fully overlap}: An overlap where the \textit{overlappee}\footref{ft:overlappee} continues speaking after the \textit{overlapper}\footref{ft:overlapper} stops speaking.
\end{itemize}

When the overlapper interrupts the overlappee and successfully takes the turn, this is termed a ``successful interjection''. Fully overlaps can result in either unsuccessful interjections or backchannels. These outcomes depend on the speaker's intention to take the turn. 
If no such intention exists, the event is a backchannel. 
We categorize backchannels into two types: 
\begin{itemize}
    \item \textbf{Emotive backchannels}: Express the listener's emotional valence, such as ``wow'' to convey surprise.
    \item \textbf{Cognitive backchannels}: Reflect the listener's cognitive state or thought processes like ``I see'' to indicate comprehension.
\end{itemize}

\noindent
Classifying backchannels requires contextual analysis. We also propose a pipeline for effective backchannel classification.

\section{Methodology}
\subsection{Overall Pipeline}

We designed a pipeline to construct a dialog dataset exclusively from two-channel speech-only datasets. The resulting dataset includes timestamps, laughter labels, response labels, emotions, sentiments for each response, and personality labels for each speaker. We began by preprocessing the speech data to construct dialog structures. Subsequently, we predicted personality traits via Large Language Models (LLMs). This prediction system integrates multiple features, including textual information (content, emotion, sentiment), acoustic features (laughter), and conversational traits (e.g., number of turns).

\subsection{Dataset Preprocessing}\label{sec:dialog_generation}

We preprocess two-channel speech data to construct dialog datasets. Two-channel speech data comprises separate channels, with each channel containing speech only from one of two speakers. 
We detect and annotate overlapping speech segments, which naturally occur in conversational settings.
The resulting dialog dataset includes timestamps, response labels (turns, backchannels, or interjections), and text-based annotations for emotion and sentiment for each response. The preprocessing procedure consists of six main steps.

\noindent
\textbf{Step 1. Speech Transcription with Word-Level Timestamps}\\
We transcribed the speech and obtained word-level timestamps using Whisper Turbo\footnote{Whisper Turbo: \url{https://github.com/openai/whisper}}~\cite{Whisper}. Subsequently, we removed silent intervals from each timestamp to refine the data.

\noindent
\textbf{Step 2. Laughter Event Annotation}\\
Laughter is a key feature for determining personality traits. Therefore, we integrated laugh labels into the system. We obtained laughter timestamps using a laughter detector\footnote{\url{https://github.com/jrgillick/laughter-detection}}~\cite{Laughter}. To enhance transcription, we added three laugh tokens based on their speech containment: \texttt{[Laughter]} for isolated vocal bursts, and \texttt{[StartLaugh]} and \texttt{[EndLaugh]} for speech-laughter co-occurrences. For example, in Fig.~\ref{fig:overall}(2), the timestamp of the first laugh includes two words, ``come'' and ``on'', resulting in ``[StartLaugh] come on [EndLaugh]''. The second laugh contains no text and is labeled as ``[Laughter]''.

\noindent
\textbf{Step 3. Response Boundary Construction}\\
We constructed sentences by concatenating words using word-level timestamps. For a single speaker, we merged successive words when the inter-word gap was below a predefined threshold (Fig.\ref{fig:overall}(c)). We set this threshold at 700ms, following~\cite{Levinson2015}, which reported language generation latencies exceeding 600ms. We defined response boundaries using silent intervals rather than contextual cues. Prior studies~\cite{Stivers2009UniversalsAC,Emanuel2007Sequence} indicate that speakers often signal turn transitions with silence, and we consider turn-taking frequency a critical feature in personality detection.

\noindent
\textbf{Step 4. Overlap Identification}\\
After constructing sentences with timestamps for both speakers, we detected overlaps. We identified two types of overlaps: ``Partial overlap'' and ``Fully overlap'' (refer to Sec.~\ref{sec:definition} for definitions). Overlap labels were excluded if the duration of overlap was less than 700 ms. We categorized responses with ``Fully-overlap'' labels as potential backchannels. Among responses labeled as ``Partial overlap'', we classified the latter response (overlapper) as a ``successful interjection''.

\noindent
\textbf{Step 5. Backchannel Classification via LLMs}\\
We classified the possible backchannels into emotive backchannels, cognitive backchannels, or unsuccessful interjections using Large Language Models (LLMs). Backchannels vary in nature and depend heavily on context~\cite{Schegloff1982,Gardner2001,Fujimoto2007}, and they are not always limited to a predefined set of words, as they can be longer or more complex~\cite{Mereu2024BackchannelsAN}. 

To address this, we employed classifiers that consider chat history. Specifically, we used GPT-4o\footnote{gpt-4o-2024-11-20: \url{https://openai.com/api/}\label{ft:4o}} to analyze and classify possible backchannels effectively.
We included the following features in the prompt for classification: 
(1) an explanation of the backchannel classes---``emotive'', ``cognitive'', and ``not backchannel''; 
(2) the chat history, including both past and future responses; and 
(3) the target text with its corresponding position in the overlappee's speech. 
The full version of the prompt is available on our project page\footref{ft:project_page}. Responses labeled as ``not backchannel'' were classified as ``unsuccessful interjection.''

\noindent
\textbf{Step 6. Emotion and Sentiment Classification of Responses}\\
Emotion and sentiment are critical features for personality prediction. We obtained those labels using text-based classifiers: the emotion classifier\footnote{\url{https://huggingface.co/j-hartmann/emotion-english-distilroberta-base}} and the sentiment classifier\footnote{\url{https://huggingface.co/cardiffnlp/twitter-roberta-base-sentiment-latest}}.
  
\subsection{Predicting Personality Traits}
We predict speaker personalities from dialogue data generated in Sec.\ref{sec:dialog_generation}. We employ an LLM (GPT-4o\footref{ft:4o}) to estimate the Big Five Personality Traits---\textit{openness}, \textit{conscientiousness}, \textit{extraversion}, \textit{agreeableness}, and \textit{neuroticism} (see Sec.\ref{sec:bfi}). We instruct the LLM to classify each speaker using five labels per personality: ``highly aligned'', ``aligned'', ``neutral'', ``opposed'', and ``highly opposed''. We assign the ``opposed'' label when a speaker exhibits the inverse of a trait (e.g., closeness instead of openness). We represent each speaker with attributes across four categories: \textit{Emotion}, \textit{Sentiment}, \textit{Basics}, and \textit{Samples}.

We convert those speaker's attributes into textual format and incorporate them into the LLM prompt. The full prompt comprises the task definition, explanations of the personality dimensions, and the speaker's attributes across the four categories. The complete prompt is available on our project page\footref{ft:project_page}. The following section details our integration of the four categories of speaker attributes into the prompt.

\begin{itemize}
\item \textbf{Emotion/Sentiment} \\
We employ emotion and sentiment labels derived from the dialog dataset. We calculate the percentage of each class (e.g., ``Anger'') within each category (e.g., ``Emotion'') for all responses associated with each speaker. We then incorporate these percentages into the LLM prompt (see Fig.~\ref{fig:prompt_emotion}).

\begin{figure}[!htb]
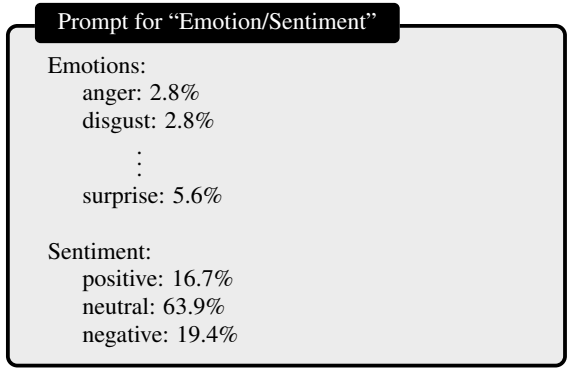

\begin{center}
\begin{AIboxSmall}{Prompt for ``Emotion/Sentiment''}
Emotions:\\
\textcolor{gray!15}{.}~\hspace{0.2cm} anger: 2.8\% \\
\textcolor{gray!15}{.}~\hspace{0.2cm} disgust: 2.8\% \\
$\textcolor{gray!15}{.}~\hspace{1.0cm} \vdots$\\
\textcolor{gray!15}{.}~\hspace{0.2cm} surprise: 5.6\% \\
\\
Sentiment:\\
\textcolor{gray!15}{.}~\hspace{0.2cm}   positive: 16.7\% \\
\textcolor{gray!15}{.}~\hspace{0.2cm}   neutral: 63.9\% \\
\textcolor{gray!15}{.}~\hspace{0.2cm}  negative: 19.4\% \\
\end{AIboxSmall}
\end{center}
\vspace{-0.7em}
\caption{
Example Prompt for ``Emotion/Sentiment''
}
\label{fig:prompt_emotion}
\end{figure}

\item \textbf{Basics} \\
We analyze the speaker's conversational attributes based on timestamps, laughter, backchannel, and interjection labels. Specifically, we incorporate the following attributes:
\begin{itemize}
  \item \textit{Number of Turns}: The number of turns, defined as responses separated by 700 ms, excluding backchannels and unsuccessful interjections.
  \item \textit{Speaking Duration}: The average speech duration per turn.
  \item \textit{Average Laughters}: The number of laughter occurrences per minute of speech.
  \item \textit{Average Emotive/Cognitive Backchannels}: The number of emotive or cognitive backchannels per minute during another speaker's speech.
  \item \textit{Number of Interjections}: The total count of successful and unsuccessful interjections within 12 minutes.
\end{itemize}

Although these attributes are interpretable, we observed that LLMs struggle to assess whether these values are high or low. To address this, we classified these values into five relative groups. We computed the mean and interquartile range (IQR) across all speakers, and defined the relative groups as follows:
\begin{itemize}
\item \textit{Normal} if $|\mathrm{value}| < \mathrm{mean} + 0.8 \times \mathrm{IQR}$
\item \textit{Many/Few} if $|\mathrm{value}| > \mathrm{mean} + 0.8 \times \mathrm{IQR}$
\item \textit{Very Many/Few} if $|\mathrm{value}| > \mathrm{mean} + 1.2 \times \mathrm{IQR}$
\end{itemize}

We then incorporate these five relative groups of the analyzed features into the prompt (see Fig.~\ref{fig:prompt_basics}).

\begin{figure}[!htb]
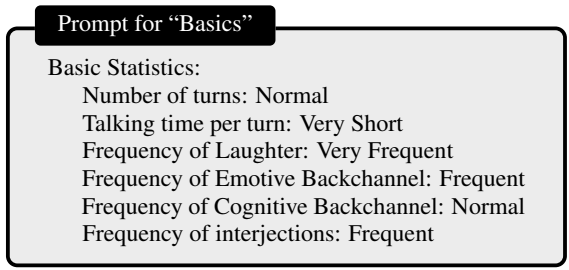

\begin{center}
\begin{AIboxSmall}{Prompt for ``Basics''}
Basic Statistics:\\
\textcolor{gray!15}{.}~\hspace{0.2cm} Number of turns: Normal\\
\textcolor{gray!15}{.}~\hspace{0.2cm}   Talking time per turn: Very Short\\
 \textcolor{gray!15}{.}~\hspace{0.2cm}  Frequency of Laughter: Very Frequent\\
\textcolor{gray!15}{.}~\hspace{0.2cm}   Frequency of Emotive Backchannel: Frequent\\
\textcolor{gray!15}{.}~\hspace{0.2cm}   Frequency of Cognitive Backchannel: Normal\\
  \textcolor{gray!15}{.}~\hspace{0.2cm} Frequency of interjections: Frequent\\
\end{AIboxSmall}
\end{center}
\vspace{-0.7em}
\caption{
Example Prompt for ``Basics''
}
\label{fig:prompt_basics}
\end{figure}

\item \textbf{Samples} \\
We include sample responses in the LLM prompt to analyze each speaker's linguistic behavior. We randomly select 20 turns exceeding 2~seconds in duration (see Fig.~\ref{fig:prompt_samples}).

\begin{figure}[!htb]
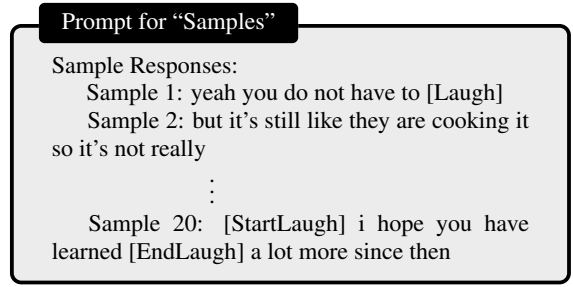

\begin{center}
\begin{AIboxSmall}{Prompt for ``Samples''}
Sample Responses:\\
\textcolor{gray!15}{.}~\hspace{0.2cm} Sample 1: yeah you do not have to [Laugh]\\
\textcolor{gray!15}{.}~\hspace{0.2cm} Sample 2: but it's still like they are cooking it so it's not really\\
$\textcolor{gray!15}{.}~\hspace{1.9cm} \vdots$\\
\textcolor{gray!15}{.}~\hspace{-0.1cm} Sample 20: [StartLaugh] i hope you have learned [EndLaugh] a lot more since then\\
\end{AIboxSmall}
\end{center}
\vspace{-0.7em}
\caption{
Example Prompt for ``Samples''
}
\label{fig:prompt_samples}
\end{figure}

\end{itemize}

\section{Experiment Setups}

\subsection{Dataset}
We used the Fisher dataset~\cite{Fisher}, a large-scale collection of telephonic speech recordings. Each sample contains a 12-minute phone conversation between two speakers, with audio captured on separate channels.
For this study, we evaluated our pipeline on a subset of the Fisher dataset~\cite{Fisher} (folder 000), consisting of 95 conversations and 190 speakers.
    
\subsection{Subjective Evaluation Experiments}
We conducted two MUSHRA tests to assess the alignment between human judgments and prediction outcomes. In the first experiment, participants predicted trends in conversational characteristics based on personality traits (e.g., between ``Frequent Laughters'' and ``Openness''). In the second experiment, evaluators rated alignment for each personality dimension using a 90-second speech dialogue. We recruited 20 English-speaking participants, each receiving comprehensive explanations of the personality traits. A sample page is available on our page\footref{ft:project_page}.

\subsection{Baselines}

We employed three baseline systems to predict personality traits from text responses. The first baseline, ``LM''\footnote{\url{https://huggingface.co/KevSun/Personality\_LM}}~\cite{PDContinuous}, was trained on Reddit comments and posts and outputs a probability distribution over five personality dimensions. The remaining baselines, BERT (``BERT'')\footnote{\url{https://huggingface.co/Minej/bert-base-personality}}~\cite{BERT} and MiniLM (``MiniLM'')\footnote{\url{https://huggingface.co/microsoft/MiniLM-L12-H384-uncased}}~\cite{MiniLM}, predict personality labels in a multi-label setting using distinct architectures. We computed personality probabilities for all responses and averaged them per speaker to derive alignment scores for each personality label.

\section{Experiments and Results}


        
\subsection{Personality Prediction}
We conducted experiments to compare our system with baseline methods. We classified speaker personality using GPT-4o\footref{ft:4o}, which predicts the alignment between speaker attributes and each of the five personalities. Classification labels are ``highly aligned'' (100), ``aligned'' (50), ``neutral'' (0), ``opposed'' (-50), and ``highly opposed'' (-100). We averaged five responses to yield reliable alignment scores, with the alignment values enclosed in curly braces.

We employed two subjective metrics to assess the alignment between human evaluations and model predictions. The first metric quantifies how well model predictions conformed to the expected trends in conversation characteristics for each personality. Our subjective evaluation confirmed these trends, as shown in Table~\ref{table:trend} (with 100 denoting ``very aligned'' and -100 denoting ``very opposed''). Consequently, we identified several attributes strongly associated with each personality. For instance, extraversion---a talkative or energetic personality---aligned with frequent turn-taking (74), while opposed to frequent fearful responses (-59).

\begin{table}[!h]
\caption{
The expected trend between speaker attributes and personalities. Blue and red denote positive and negative correlations, respectively. Strong correlations are highlighted in bold. Column names indicate abbreviated personality labels.
}
\label{table:trend}
\centering
\scalebox{0.80}{
\begin{tabular}{lccccc}
\toprule
  & Ope. & Con. & Ext. & Agr. & Neu.\\
\midrule
Emotion --- Anger &\textcolor{red!60}{-44} & \textbf{\textcolor{red}{-55}} & \textcolor{red!60}{-30} & \textbf{\textcolor{red}{-70}} & \textbf{\textcolor{blue}{64}}\\
Emotion --- Disgust &\textbf{\textcolor{red}{-63}} & \textbf{\textcolor{red}{-51}} & \textcolor{red!60}{-47} & \textbf{\textcolor{red}{-77}} & \textbf{\textcolor{blue}{57}}\\
Emotion --- Fear &\textbf{\textcolor{red}{-70}} & \textbf{\textcolor{red}{-61}} & \textbf{\textcolor{red}{-59}} & \textbf{\textcolor{red}{-66}} & \textbf{\textcolor{blue}{64}}\\
Emotion --- Joy &\textbf{\textcolor{blue}{50}} & \textcolor{blue!60}{41} & \textbf{\textcolor{blue}{65}} & \textbf{\textcolor{blue}{61}} & \textcolor{red!60}{-16}\\
Emotion --- Neutral &\textcolor{blue!60}{25} & \textcolor{blue!60}{37} & \textcolor{blue!60}{17} & \textcolor{blue!60}{31} & \textcolor{red!60}{-16}\\
Emotion --- Sadness &\textbf{\textcolor{red}{-55}} & \textcolor{red!60}{-47} & \textbf{\textcolor{red}{-50}} & \textbf{\textcolor{red}{-52}} & \textbf{\textcolor{blue}{58}}\\
Emotion --- Surprise &\textcolor{blue!60}{39} & \textcolor{blue!60}{35} & \textbf{\textcolor{blue}{71}} & \textcolor{blue!60}{42} & \textcolor{blue!60}{47}\\
\addlinespace[0.1em]\hdashline\addlinespace[0.3em]
Sentiment --- Positive &\textbf{\textcolor{blue}{50}} & \textcolor{blue!60}{47} & \textbf{\textcolor{blue}{62}} & \textbf{\textcolor{blue}{63}} & \textcolor{red!60}{-49}\\
Sentiment --- Neutral &\textcolor{blue!60}{31} & \textcolor{blue!60}{42} & \textcolor{blue!60}{20} & \textcolor{blue!60}{17} & \textcolor{red!60}{-17}\\
Sentiment --- Negative &\textcolor{red!60}{-42} & \textbf{\textcolor{red}{-54}} & \textcolor{red!60}{-33} & \textbf{\textcolor{red}{-64}} & \textbf{\textcolor{blue}{56}}\\
\addlinespace[0.1em]\hdashline\addlinespace[0.3em]
Basics --- Taking Turns &\textbf{\textcolor{blue}{57}} & \textbf{\textcolor{blue}{54}} & \textbf{\textcolor{blue}{74}} & \textcolor{blue!60}{47} & \textcolor{red!60}{-15}\\
Basics --- Average Time per Turn &\textbf{\textcolor{blue}{62}} & \textcolor{blue!60}{15} & \textbf{\textcolor{blue}{72}} & \textcolor{blue!60}{3} & \textcolor{red!60}{-5}\\
Basics --- Laughters &\textcolor{blue!60}{17} & \textcolor{blue!60}{12} & \textbf{\textcolor{blue}{66}} & \textbf{\textcolor{blue}{51}} & \textcolor{red!60}{-9}\\
Basics --- Emotive Backchannels &\textcolor{blue!60}{43} & \textcolor{blue!60}{48} & \textbf{\textcolor{blue}{73}} & \textbf{\textcolor{blue}{68}} & \textcolor{blue!60}{1}\\
Basics --- Cognitive Backchannels &\textbf{\textcolor{blue}{51}} & \textbf{\textcolor{blue}{61}} & \textbf{\textcolor{blue}{68}} & \textbf{\textcolor{blue}{77}} & \textcolor{red!60}{-15}\\
Basics --- Interjections &\textcolor{blue!60}{1} & \textcolor{red!60}{-30} & \textbf{\textcolor{blue}{58}} & \textcolor{red!60}{-39} & \textcolor{blue!60}{6}\\
\bottomrule
\end{tabular}
}
\caption*{This table is derived solely from a human survey}
\end{table}
\vspace{-1.2em}

We quantified the alignment between speaker conversation attributes and predicted personality scores. For each personality label, we computed a weighted sum of correlation coefficients between personality traits and raw speaker attributes (prior to categorization), where higher values indicate stronger positive correlations. We derive the weights by normalizing the values in Table~\ref{table:trend}. Here, we multiplied coefficients corresponding to negative trends by $-1$. Table~\ref{table:score_trend} summarizes these scores. Our model outperforms baseline models in most metrics; however, we observed a lower score in ``Conscientiousness''. Our analysis reveals that the correlation for the ``Emotion'' attribute deviates from expected trends, with values of $-0.166$, $0.0176$, and $0.079$ for ``surprise'', ``angry'', and ``sadness'', respectively. We attribute these discrepancies to the model's reliance on sample responses for determining responsibility, which emotional cues may not adequately capture. 


\begin{table}[!h]
\caption{
Alignment of human judgments on conversation trait trends. 
Column names denote abbreviated personality labels.
}
\label{table:score_trend}
\centering
\scalebox{0.80}{
\begin{tabular}{lcccccc}
\toprule
\hspace{1em}($\uparrow$) & Open. & Cons. & Extr. & Agre. & Neur. & Avg.\\
\midrule
Ours &\textbf{0.167} & \textbf{0.061} & \textbf{0.174} & \textbf{0.210} & \textbf{0.315} & \textbf{0.186}\\
\addlinespace[0.1em]\hdashline\addlinespace[0.3em]
BERT &0.053 & -0.04 & 0.036 & -0.015 & -0.087 & -0.011\\
LM &0.030 & -0.011 & -0.021 & 0.044 & 0.144 & 0.037\\
MiniLM &0.060 & 0.035 & 0.010 & 0.123 & 0.122 & 0.070\\
\bottomrule
\end{tabular}
}
\end{table}
In our second evaluation, we computed the similarity between predicted personality scores and human-annotated labels obtained from a listening test. Table~\ref{table:score_label} reports the correlation values for each personality trait and the average cosine similarity across all samples. Our model consistently outperforms the other baselines, particularly when we interpret personality labels as vector dimensions (Cosine Similarity).

\begin{table}[!h]
\caption{
Similarities between human labels and prediction results. Column names denote abbreviated personality labels.
}
\label{table:score_label}
\centering
\scalebox{0.73}{
\begin{tabular}{lccccccc}
\toprule
 & \multicolumn{6}{c}{Correlation from Ground-Truth Labels (Corr.) ($\uparrow$)} & \multicolumn{1}{c}{Cosine}\\
\cmidrule(lr){2-7}\cmidrule(lr){8-8}
  & Open. & Cons. & Extr. & Agre. & Neur. & Avg. & Similarity\\
\midrule
Ours &\textbf{0.169} & \textbf{0.198} & \textbf{0.285} & \textbf{0.144} & \textbf{0.117} & \textbf{0.183} & \textbf{0.503}\\
\addlinespace[0.1em]\hdashline\addlinespace[0.3em]
BERT &0.095 & 0.044 & 0.194 & 0.058 & 0.089 & 0.096 & -0.543\\
LM &0.023 & -0.09 & -0.018 & 0.030 & -0.027 & -0.016 & -\\
MiniLM &-0.122 & -0.064 & 0.014 & -0.112 & 0.021 & -0.053 & 0.148\\
\bottomrule
\end{tabular}
}
\end{table}

%

\subsection{Ablation Study}

We conducted an ablation study on character prediction by varying speaker attributes to assess the impact of the attribute categories such as Emotion/Sentiment, Basics, and Samples. We collected five responses for each experimental condition. Table~\ref{table:ablation} shows that the full prompt consistently outperforms other conditions across all metrics, indicating improved alignment with human judgment and enhanced correlation with expected trends. In particular, compared with the ``Samples''-only condition, the higher ``Trend'' score confirms that we effectively direct the prediction focus to the desired attributes.

\begin{table}[!h]
\caption{
An ablation study of speaker characteristics. ``O'' and ``-'' denote the inclusion and exclusion of features, respectively. Averaged scores in Table~\ref{table:score_trend} and Table~\ref{table:score_label} are denoted as ``Trend'', ``Corr.'', and ``Cosine'', respectively.
}
\label{table:ablation}
\centering
\scalebox{0.80}{
\begin{tabular}{ccccccc}
\toprule
 \multicolumn{4}{c}{Conditions} & \multicolumn{3}{c}{Scores}\\
\cmidrule(lr){1-4}\cmidrule(lr){5-7}
 Samples & Basics & Emotion & Sentiment & Trend & Corr. & Cosine\\
\midrule
\textcolor{black}{O} & \textcolor{gray}{-} & \textcolor{gray}{-} & \textcolor{gray}{-} & 0.108 & 0.167 & 0.551\\
\textcolor{gray}{-} & \textcolor{black}{O} & \textcolor{gray}{-} & \textcolor{gray}{-} & 0.051 & 0.009 & 0.524\\
\textcolor{gray}{-} & \textcolor{gray}{-} & \textcolor{black}{O} & \textcolor{black}{O} & \textbf{0.277} & 0.064 & -0.203\\
\textcolor{black}{O} & \textcolor{black}{O} & \textcolor{gray}{-} & \textcolor{gray}{-} & 0.102 & 0.135 & \textbf{0.603}\\
\textcolor{black}{O} & \textcolor{gray}{-} & \textcolor{black}{O} & \textcolor{black}{O} & 0.173 & 0.170 & 0.407\\
\textcolor{gray}{-} & \textcolor{black}{O} & \textcolor{black}{O} & \textcolor{black}{O} & 0.210 & 0.089 & 0.198\\
\textcolor{black}{O} & \textcolor{black}{O} & \textcolor{black}{O} & \textcolor{black}{O} & 0.186 & \textbf{0.183} & 0.503\\
\bottomrule
\end{tabular}
}
\end{table}
\section{Conclusion}
We propose a pipeline for constructing a dialogue dataset with extensive labels and for predicting personality labels from speech-only data. Our analysis shows that our prediction model aligns more closely with human judgments than existing models. Future work will investigate synthetic datasets with personality labels, motivated by evidences that LLMs can generate dialogue based on personality traits~\cite{Wanqi2020ResponseGB,Chen2023}, and will develop conversational agents conditioned on personality labels.


\section{Acknowledgements}
Research is supported by (1) Shenzhen Science and Technology Program (Shenzhen Key Laboratory, Grant No. ZDSYS20230626091302006), (2) Shenzhen Science and Technology Research Fund (Fundamental Research Key Project, Grant No. JCYJ20220818103001002), and (3) Program for Guangdong Introducing Innovative and Enterpreneurial Teams, Grant No. 2023ZT10X044.

\newpage

\bibliographystyle{IEEEtran}
\bibliography{mybib_nourl}

\begin{thebibliography}{10}
\providecommand{\url}[1]{#1}
\csname url@samestyle\endcsname
\providecommand{\newblock}{\relax}
\providecommand{\bibinfo}[2]{#2}
\providecommand{\BIBentrySTDinterwordspacing}{\spaceskip=0pt\relax}
\providecommand{\BIBentryALTinterwordstretchfactor}{4}
\providecommand{\BIBentryALTinterwordspacing}{\spaceskip=\fontdimen2\font plus
\BIBentryALTinterwordstretchfactor\fontdimen3\font minus \fontdimen4\font\relax}
\providecommand{\BIBforeignlanguage}[2]{{%
\expandafter\ifx\csname l@#1\endcsname\relax
\typeout{** WARNING: IEEEtran.bst: No hyphenation pattern has been}%
\typeout{** loaded for the language `#1'. Using the pattern for}%
\typeout{** the default language instead.}%
\else
\language=\csname l@#1\endcsname
\fi
#2}}
\providecommand{\BIBdecl}{\relax}
\BIBdecl

\bibitem{SurveySDM}
S.~Ji, Y.~Chen, M.~Fang, J.~Zuo, J.~Lu, H.~Wang, Z.~Jiang, L.~Zhou, S.~Liu, X.~Cheng \emph{et~al.}, ``Wavchat: A survey of spoken dialogue models,'' \emph{arXiv preprint arXiv:2411.13577}, 2024.

\bibitem{SpeechGPT}
D.~Zhang, S.~Li, X.~Zhang, J.~Zhan, P.~Wang, Y.~Zhou, and X.~Qiu, ``Speechgpt: Empowering large language models with intrinsic cross-modal conversational abilities,'' in \emph{Conference on Empirical Methods in Natural Language Processing}, 2023.

\bibitem{QAandSC}
E.~Nachmani, A.~Levkovitch, R.~Hirsch, J.~Salazar, C.~Asawaroengchai, S.~Mariooryad, E.~Rivlin, R.~Skerry-Ryan, and M.~T. Ramanovich, ``Spoken question answering and speech continuation using spectrogram-powered {LLM},'' in \emph{The Twelfth International Conference on Learning Representations}, 2024.

\bibitem{Llamaomni}
Q.~Fang, S.~Guo, Y.~Zhou, Z.~Ma, S.~Zhang, and Y.~Feng, ``Llama-omni: Seamless speech interaction with large language models,'' 2024.

\bibitem{dGSLM}
T.~A. Nguyen, E.~Kharitonov, J.~Copet, Y.~Adi, W.-N. Hsu, A.~Elkahky, P.~Tomasello, R.~Algayres, B.~Sagot, A.~Mohamed, and E.~Dupoux, ``Generative spoken dialogue language modeling,'' 2022.

\bibitem{Moshi}
A.~D'efossez, L.~Mazar'e, M.~Orsini, A.~Royer, P.~P'erez, H.~J'egou, E.~Grave, and N.~Zeghidour, ``Moshi: a speech-text foundation model for real-time dialogue,'' \emph{ArXiv}, vol. abs/2410.00037, 2024.

\bibitem{DialogueScheme}
P.~Wang, S.~Lu, Y.~Tang, S.~Yan, W.~Xia, and Y.~Xiong, ``A full-duplex speech dialogue scheme based on large language model,'' in \emph{The Thirty-eighth Annual Conference on Neural Information Processing Systems}, 2024.

\bibitem{Pipek2007OnBI}
V.~Pipek, ``On backchannels in english conversation,'' 2007.

\bibitem{Bavelas2000}
J.~Bavelas, L.~Coates, and T.~Johnson, ``Listeners as co-narrators,'' \emph{Journal of personality and social psychology}, vol.~79, pp. 941--52, 12 2000.

\bibitem{Tolins2015}
J.~Tolins and J.~Fox~Tree, ``Overhearers use addressee backchannels in dialog comprehension,'' \emph{Cognitive science}, vol.~40, 09 2015.

\bibitem{Sanfiel2014}
M.~Soto-Sanfiel, L.~Aymerich-Franch, E.~Romero, M.~Soto, M.~Soto, and M.~Sanfiel, ``Personality in interaction: how the big five relate to the reception of interactive narratives,'' \emph{COMMUNICATION \& SOCIETY / COMUNICACIÓN Y SOCIEDAD}, 06 2014.

\bibitem{Mairesse2007}
F.~Mairesse, M.~A. Walker, M.~R. Mehl, and R.~K. Moore, ``Using linguistic cues for the automatic recognition of personality in conversation and text,'' \emph{Journal of Artificial Intelligence Research}, vol.~30, pp. 457--500, 2007.

\bibitem{Park2024}
Y.-H. Park, W.~Liermann, Y.-S. Choi, S.~Kim, J.-U. Bang, S.~Yun, and K.~Lee, ``Backchannel prediction, based on who, when and what,'' 09 2024, pp. 3570--3574.

\bibitem{Han2010}
H.~Li, Y.~Cui, and Z.~Wang, ``Backchannel responses and enjoyment of the conversation: The more does not necessarily mean the better,'' \emph{International Journal of Psychological Studies}, vol.~2, 05 2010.

\bibitem{1987129}
S.~K. MAYNARD, ``On back-channel behavior in japanese and english casual conversation,'' \emph{Gengokenkyu}, vol. 1987, no.~91, pp. 129--130, 1987.

\bibitem{PDContinuous}
R.~Wang and K.~Sun, ``Continuous output personality detection models via mixed strategy training,'' 06 2024.

\bibitem{PDmultiplesns}
H.~Christian, D.~Suhartono, A.~Chowanda, and K.~Zamli, ``Text based personality prediction from multiple social media data sources using pre-trained language model and model averaging,'' \emph{Journal of Big Data}, vol.~8, 05 2021.

\bibitem{PDChatGPT}
Y.~Ji, W.~Wu, H.~Zheng, Y.~Hu, X.~Chen, and L.~He, ``Is chatgpt a good personality recognizer? a preliminary study,'' 07 2023.

\bibitem{PDAPP}
M.~Ramezani, M.-R. Feizi-Derakhshi, M.~A. Balafar, M.~Asgari-Chenaghlu, A.~R.~F. Derakhshi, N.~Nikzad-Khasmakhi, M.~Ranjbar-Khadivi, Z.~Jahanbakhsh-Nagadeh, E.~Zafarani-Moattar, and T.~Rahkar-Farshi, ``Automatic personality prediction: an enhanced method using ensemble modeling,'' \emph{Neural Computing and Applications}, vol.~34, pp. 18\,369 -- 18\,389, 2020.

\bibitem{PDbottomtop}
Y.~Mehta, S.~Fatehi, A.~Kazameini, C.~Stachl, E.~Cambria, and S.~Eetemadi, ``Bottom-up and top-down: Predicting personality with psycholinguistic and language model features,'' in \emph{2020 IEEE International Conference on Data Mining (ICDM)}, 2020, pp. 1184--1189.

\bibitem{PDgraywolf}
H.~Lin, C.~Wang, and Q.~Hao, ``A novel personality detection method based on high-dimensional psycholinguistic features and improved distributed gray wolf optimizer for feature selection,'' \emph{Information Processing \& Management}, vol.~60, no.~2, p. 103217, 2023.

\bibitem{PDDyadic}
Q.~Liu, ``Modeling dyadic conversations for personality inference,'' 09 2020.

\bibitem{BFI}
J.~M. Digman, ``Personality structure: Emergence of the five-factor model,'' \emph{Annual Review of Psychology}, vol.~41, no.~1, pp. 417--440, 1990.

\bibitem{TurntakingICLR}
Anonymous, ``Talking turns: Benchmarking audio foundation models on turn-taking dynamics,'' in \emph{Submitted to The Thirteenth International Conference on Learning Representations}, 2024, under review.

\bibitem{Duncan1972}
S.~Duncan, ``Some signals and rules for taking speaking turns in conversations,'' \emph{Journal of Personality and Social Psychology}, vol.~23, no.~2, pp. 283--292, 1972.

\bibitem{Ameka2006}
F.~K. Ameka, ``Interjections,'' in \emph{Encyclopedia of Language \& Linguistics}, K.~Brown, Ed.\hskip 1em plus 0.5em minus 0.4em\relax Elsevier, 2006, pp. 743--746.

\bibitem{SKANTZE2021101178}
G.~Skantze, ``Turn-taking in conversational systems and human-robot interaction: A review,'' \emph{Computer Speech \& Language}, vol.~67, p. 101178, 2021.

\bibitem{Whisper}
A.~Radford, J.~W. Kim, T.~Xu, G.~Brockman, C.~McLeavey, and I.~Sutskever, ``Robust speech recognition via large-scale weak supervision,'' in \emph{International conference on machine learning}.\hskip 1em plus 0.5em minus 0.4em\relax PMLR, 2023, pp. 28\,492--28\,518.

\bibitem{Laughter}
J.~Gillick, W.~Deng, K.~Ryokai, and D.~Bamman, ``Robust laughter detection in noisy environments,'' 08 2021, pp. 2481--2485.

\bibitem{Levinson2015}
S.~C. Levinson and F.~Torreira, ``Timing in turn-taking and its implications for processing models of language,'' \emph{Frontiers in Psychology}, vol.~6, p. 731, 2015.

\bibitem{Stivers2009UniversalsAC}
T.~Stivers, N.~J. Enfield, P.~Brown, C.~Englert, M.~Hayashi, T.~Heinemann, G.~Hoymann, F.~Rossano, J.~P.~D. Ruiter, K.-E. Yoon, S.~C. Levinson, P.~Kay, and K.~Y, ``Universals and cultural variation in turn-taking in conversation,'' \emph{Proceedings of the National Academy of Sciences}, vol. 106, pp. 10\,587 -- 10\,592, 2009.

\bibitem{Emanuel2007Sequence}
E.~Schegloff, ``Sequence organization in interaction: A primer in conversation analysis,'' \emph{Sequence Organization in Interaction: A Primer in Conversation Analysis I}, vol.~1, pp. 1--300, 01 2007.

\bibitem{Schegloff1982}
------, \emph{Discourse as an interactional achievement: Some uses of `uh huh' and other things that come between sentences}, 01 1982, pp. 71--93.

\bibitem{Gardner2001}
R.~Gardner, \emph{When listeners talk: Response tokens and listener stance}.\hskip 1em plus 0.5em minus 0.4em\relax John Benjamins Publishing Company, 2001.

\bibitem{Fujimoto2007}
D.~Fujimoto, ``Listener responses in interaction: A case for abandoning the term, backchannel,'' \emph{Journal of Osaka Jogakuin 2year College}, vol.~37, 01 2007.

\bibitem{Mereu2024BackchannelsAN}
D.~Mereu, F.~Cangemi, and M.~Grice, ``Backchannels are not always very short utterances. the case of italian multi-unit backchannels,'' \emph{Journal of Pragmatics}, 2024.

\bibitem{Fisher}
C.~Cieri, D.~Miller, and K.~Walker, ``Fisher english training speech parts 1 and 2,'' 2004, accessed: 2025-01-09.

\bibitem{BERT}
J.~Devlin, M.~Chang, K.~Lee, and K.~Toutanova, ``{BERT:} pre-training of deep bidirectional transformers for language understanding,'' \emph{CoRR}, vol. abs/1810.04805, 2018.

\bibitem{MiniLM}
W.~Wang, F.~Wei, L.~Dong, H.~Bao, N.~Yang, and M.~Zhou, ``Minilm: Deep self-attention distillation for task-agnostic compression of pre-trained transformers,'' 2020.

\bibitem{Wanqi2020ResponseGB}
W.~Wanqi and T.~Sakai, ``Response generation based on the big five personality traits,'' 2020.

\bibitem{Chen2023}
E.~Chen, ``Generate labeled training data using prompt programming and gpt-3. an example of big five personality classification,'' 03 2023.

\end{thebibliography}

\end{document}